\documentstyle[epsf,preprint,aps]{revtex}

\begin{document}

\preprint{UTCCP-P-36, June 1998}  

\title{Non-perturbative determination of anisotropy coefficients\\
in lattice gauge theories}

\author{S.\ Ejiri, Y.\ Iwasaki, and K.\ Kanaya
} 

\address{
Center for Computational Physics, University of Tsukuba,\\
Tsukuba, Ibaraki 305 , Japan}

\maketitle

\begin{abstract}
We propose a new non-perturbative method 
to compute derivatives of gauge coupling constants 
with respect to anisotropic lattice spacings (anisotropy coefficients),
which are required in an evaluation of thermodynamic quantities from 
numerical simulations on the lattice.
Our method is based on a precise measurement of the finite temperature 
deconfining transition curve in the lattice coupling parameter space 
extended to anisotropic lattices 
by applying the spectral density method.
We test the method for the cases of $SU(2)$ and $SU(3)$ gauge theories
at the deconfining transition point on lattices with the lattice 
size in the time direction  $N_t=4$ -- 6. 
In both cases, there is a clear discrepancy between our results
and perturbative values.
A longstanding problem, 
when one uses the perturbative anisotropy coefficients, 
is a non-vanishing pressure gap at the deconfining transition point 
in the $SU(3)$ gauge theory.
Using our non-perturbative anisotropy coefficients, we find that
this problem is completely resolved:  
we obtain
$\Delta p/T^4 = 0.001(15)$ and $-0.003(17)$ on $N_t=4$ and 6 lattices,
respectively.
\end{abstract}

\vspace{10mm}

\section{Introduction}
\label{sec:intro}

In order to study the nature of the quark-gluon plasma in 
heavy ion collisions and in the early Universe, 
it is important to evaluate 
the energy density $\epsilon$ and the pressure $p$
near the transition temperature of the deconfining phase transition. 
These quantities are defined by derivatives of the partition function 
in terms of the temperature $T$ 
and the physical volume $V$ of the system
\begin{eqnarray}
\epsilon = - \frac{1}{V} \frac{\partial \ln Z}{\partial T^{-1}}, 
\hspace{5mm} 
 p       = T \frac{\partial \ln Z}{\partial V}. 
\label{eqn:ep}
\end{eqnarray}
The lattice formulation of QCD provides us with a non-perturbative way 
to compute these quantities by numerical simulations.
On a lattice with a size $N_s^3\times N_t$, $V$ and $T$ are given by 
$V = (N_s a_s)^3$ and $T = 1 / (N_t a_t)$, 
with $a_s$ and $a_t$ the lattice spacings in spatial and 
temporal directions.
Because $N_s$ and $N_t$ are discrete parameters, 
the partial differentiations in (\ref{eqn:ep}) are performed 
by varying $a_s$ and $a_t$ independently on anisotropic lattices.

The anisotropy on a lattice is realized by introducing 
different coupling parameters in temporal and spatial directions.
For an $SU(N_c)$ gauge theory, the standard plaquette action 
on an anisotropic lattice is given by
\begin{eqnarray}
 S = -\beta_s \sum_{x,\, i<j \ne 4} P_{ij}(x) 
     -\beta_t \sum_{x,\, i \ne 4}   P_{i4}(x),
\end{eqnarray}
where
$ P_{\mu \nu}(x) = \frac{1}{N_{c}} {\rm Re \ Tr} 
\{ U_{\mu}(x) U_{\nu}(x+\hat{\mu})
   U^{\dagger}_{\mu}(x+\hat{\nu}) U^{\dagger}_{\nu}(x) \}$
is the plaquette in the $(\mu,\nu)$ plane.
With this action, 
the energy density and pressure are given by \cite{satz,karsch}
\begin{eqnarray}
\epsilon &=& - \frac{3 N_t^{4} T^{4}}{\xi^3} 
  \left\{ \left(a_t \frac{\partial \beta_s}{\partial a_t} 
 -\xi \frac{\partial \beta_s}{\partial \xi}\right) 
 (\langle P_s \rangle - \langle P \rangle_0) + 
  \left(a_t \frac{\partial \beta_t}{\partial a_t} 
 -\xi \frac{\partial \beta_t}{\partial \xi}\right) 
 (\langle P_t \rangle - \langle P \rangle_0) \right\}, \label{enrg} \\
p &=& \frac{N_t^{4} T^{4}}{\xi^3} 
  \left\{\xi \frac{\partial \beta_s}{\partial \xi} \,
 (\langle P_s \rangle  - \langle P \rangle_0)
 +\xi \frac{\partial \beta_t}{\partial \xi} \,
 (\langle P_t \rangle - \langle P \rangle_0) \right\}, \label{prs}
\end{eqnarray}
where $\langle P_{s(t)} \rangle$ is
the space(time)-like plaquette expectation value 
and $\langle P \rangle_0$ the plaquette expectation value 
on a zero-temperature lattice.
Here, for later convenience, we have chosen $a_t$ and 
$\xi \equiv a_s/a_t$ as 
independent variables to vary the lattice spacings, 
instead of $a_s$ and $\xi$ adopted in \cite{karsch}.

In order to compute $\epsilon$ and $p$ from eqs.~(\ref{enrg}) 
and (\ref{prs}) using numerical results from simulations, 
the values for the derivatives of gauge coupling constants
with respect to the anisotropic lattice spacings 
\begin{eqnarray}
a_t \frac{\partial \beta_s}{\partial a_t},
\hspace{5mm}
a_t \frac{\partial \beta_t}{\partial a_t},
\hspace{5mm}
\frac{\partial \beta_s}{\partial \xi},
\hspace{5mm}
\frac{\partial \beta_t}{\partial \xi},
\end{eqnarray}
which we call the anisotropy coefficients, are required.
They can be computed from a requirement that, in the scaling region,
the effects of anisotropy in the physical observables can be absorbed 
by a renormalization of the coupling parameters.
Similar to the case of the renormalization group beta-function, 
the anisotropy coefficients do not depend on the temperature, 
because the renormalization is independent of the temperature.

The calculation of these anisotropy coefficients in the lowest order 
perturbation theory is done by Karsch\cite{karsch}.
However, the perturbative coefficients are known to lead to
pathological results such as a negative pressure 
and a non-vanishing pressure gap at the deconfining transition
in $SU(3)$ gauge theory. Therefore,
non-perturbative values of the anisotropy coefficients are required 
in order to study the thermodynamic quantities 
near the phase transition when $N_t$ is not sufficiently large.

We are interested in the values of the anisotropy coefficients for
isotropic lattices ($\beta_s=\beta_t\equiv\beta$, i.e.\ $\xi=1$)
where most simulations are performed.
In this case, we have 
$(a_t \frac{\partial \beta_s}{\partial a_t})_{\xi = 1}
= (a_t \frac{\partial \beta_t}{\partial a_t})_{\xi = 1}
= a \frac{{\rm d} \beta}{{\rm d} a} 
= 2N_{c} a \frac{{\rm d} g^{-2}}{{\rm d} a}$,
where $a \frac{{\rm d} g^{-2}}{{\rm d} a}$ is 
the beta-function at $\xi = 1$,
whose non-perturbative values are well studied both in $SU(2)$ and
$SU(3)$ gauge theories\cite{taro,engels,boyd,edwards}.
Furthermore, a combination of the remaining two anisotropy coefficients 
is known to be related to the beta-function \cite{karsch} by%
\footnote{
In \cite{karsch}, a corresponding equation is given for
$(\partial \beta_{s (t)} / \partial \xi )_{a_s: {\rm fixed}}$.
}
\begin{eqnarray}
\left(\frac{\partial \beta_s}{\partial \xi} 
+ \frac{\partial \beta_t}{\partial \xi}\right)
_{a_t : {\rm fixed},\, \xi = 1}  
= \frac{3}{2} \, a \frac{{\rm d} \beta}{{\rm d} a}. 
\label{eqn:cubicsym}
\end{eqnarray}
Therefore, only one additional input is required to determine 
the anisotropy coefficients for isotropic lattices.

A non-perturbative determination of the anisotropy coefficients 
was attempted in Refs.~\cite{burgers,fujisaki,scheideler,klassen}
using a method that we call ``the matching method'' in the following.
One first determines $\xi$ as a function of $\beta_s$ 
and $\beta_t$ by matching space-like and time-like Wilson 
loops on anisotropic lattices, 
and then numerically determines $\partial\gamma/\partial\xi$
at $\xi=1$, where $\gamma = \sqrt{\beta_t/\beta_s}$.
Interpolation of the Wilson loop data at different sizes 
or interpolation of $\xi$ at different $\gamma$'s using 
an Ansatz is required to evaluate 
$\partial\gamma/\partial\xi$ at $\xi = 1$.

Alternatively, we can evaluate a non-perturbative value of 
pressure directly from the Monte Carlo data 
by ``the integral method'' \cite{fingberg}:
Assuming homogeneity expected when the spatial lattice size is
sufficiently large,
we obtain the relation $ p = -f$, 
where $ f = - \frac{T}{V} \ln Z$ is the free energy density, 
which can be evaluated by numerically integrating the plaquette difference 
$\langle P_s\rangle + \langle P_t\rangle - 2\langle P\rangle_0$ 
in terms of $\beta$ on isotropic lattices.
The resulting value of the pressure, in turn, provides us with 
a non-perturbative estimate of 
an anisotropy coefficient \cite{engels,boyd}.
In actual numerical simulations, 
as the value of $p$ in the confining phase and near the deconfining 
transition point is quite small compared with the magnitude of errors, 
it is difficult to determine the anisotropy coefficients
near the transition point \cite{scheideler}.

In this paper, 
we propose a new method to directly compute the anisotropy coefficients 
at the deconfining transition point.
Our method is described in Sec.~\ref{sec:method}.
We test the method in the cases of $SU(2)$ gauge theory in 
Sec.~\ref{sec:su2}.
The more realistic case of $SU(3)$ gauge theory is studied in 
Sec.~\ref{sec:su3}. 
As an application of our non-perturbative anisotropy coefficients,
we study the gaps for $\epsilon$ and $p$ at the $SU(3)$ deconfining 
transition for $N_t=4$ and 6.
A summary is given in Sec.~\ref{sec:conclusions}.

\section{Method}
\label{sec:method}

Our method is based on an observation that, in the scaling region, 
the transition temperature $T_{c} = 1/\{N_t a_t (\beta_s, \beta_t)\}$ 
must be independent of the anisotropy of the lattice. 
Therefore, 
when we change the coupling constants along the transition curve 
in the $(\beta_s,\beta_t)$ plane like 
$ (\beta_s, \beta_t) \rightarrow 
(\beta_s + {\rm d} \beta_s, 
\beta_t + {\rm d} \beta_t) $ on a lattice with fixed $N_t$, 
the lattice spacing in the time direction $a_t$ does not change:
\begin{eqnarray}
{\rm d} a_{t }= \frac{\partial a_t}{\partial \beta_s} \,
{\rm d} \beta_s + \frac{\partial a_t}{\partial \beta_t} \,
{\rm d} \beta_t = 0.
\end{eqnarray}
We denote the slope of the transition curve at $\xi = 1$ by $r_t$;
\begin{eqnarray}
r_t =  
\frac{{\rm d} \beta_s}{{\rm d} \beta_t} = 
- \left(\frac{\partial a_t}{\partial \beta_t}\right)_{\xi = 1} 
\left/
  \left(\frac{\partial a_t}{\partial \beta_s}\right)_{\xi = 1}
\right. 
=  \left(\frac{\partial \beta_s}{\partial \xi}\right)_{\xi = 1} 
\left/ \left(\frac{\partial \beta_t}{\partial \xi}\right)_{\xi = 1}
\right. ,
\end{eqnarray}
where we used an identity
\begin{eqnarray}
\left( \begin{array}{cc}
\frac{\partial \beta_s}{\partial a_t} &
\frac{\partial \beta_t}{\partial a_t} \\ 
\frac{\partial \beta_s}{\partial \xi} &
\frac{\partial \beta_t}{\partial \xi} 
\end{array} \right) = 
\frac{1}{ 
 \frac{\partial \xi}{\partial \beta_t}
 \frac{\partial a_t}{\partial \beta_s}
-\frac{\partial \xi}{\partial \beta_s}
\frac{\partial a_t}{\partial \beta_t} }
\left( \begin{array}{cc}
 \frac{\partial \xi}{\partial \beta_t} &
-\frac{\partial \xi}{\partial \beta_s} \\
-\frac{\partial a_t}{\partial \beta_t} &
 \frac{\partial a_t}{\partial \beta_s}
\end{array} \right).
\end{eqnarray}
Hence, the derivatives of $\beta_s$ and $\beta_t$ 
in terms of $\xi$ are expressed as 
\begin{eqnarray}
\left(\frac{\partial \beta_s}{\partial \xi}\right)_{\xi = 1} 
&=& \frac{3 r_t}{2 ( 1 + r_t )} \,
a \frac{{\rm d} \beta}{{\rm d} a}, \nonumber \\
 \left(\frac{\partial \beta_t}{\partial \xi}\right)_{\xi = 1} 
&=& \frac{3}{2 ( 1 + r_t )} \, a \frac{{\rm d} \beta}{{\rm d} a}. 
\end{eqnarray}
Introducing the conventional notation
$\gamma = \sqrt{\beta_t/\beta_s}$ and $\beta = \sqrt{\beta_s \beta_t}$,
we obtain
\begin{eqnarray}
\left(\frac{\partial\gamma}{\partial\xi}\right)_{a_t:{\rm fixed},\,\xi=1}
= \left(\frac{\partial\gamma}{\partial\xi}\right)_{a_s:{\rm fixed},\,\xi=1}
= \frac{3}{4\beta} \, \frac{1-r_t}{1+r_t} \, a\frac{{\rm d}\beta}{{\rm d}a}.
\label{eq:dgamdxi}
\end{eqnarray}
Finally, the customarily used forms for the anisotropy coefficients 
(Karsch coefficients) \cite{karsch} are given by 
\begin{eqnarray}
c_s 
&=& \left(\frac{\partial g_s^{-2} }
    {\partial \xi}\right)_{a_s: {\rm fixed},\, \xi = 1} 
= \frac{1}{2N_{c}} \left\{ \beta + \frac{r_t - 2}{2 ( 1 + r_t)} \,
    a \frac{{\rm d} \beta}{{\rm d} a} \right\}, \nonumber \\ 
c_t 
&=& \left(\frac{\partial g_t^{-2} }
    {\partial \xi}\right)_{a_s: {\rm fixed},\, \xi = 1} 
= \frac{1}{2N_{c}} \left\{ - \beta 
    + \frac{1 - 2 r_t}{2 ( 1 + r_t)} \, 
    a \frac{{\rm d} \beta}{{\rm d} a} \right\},
\label{kc}
\end{eqnarray}
where 
$\beta_s = 2N_{c} g_s^{-2} \xi^{-1}$ and 
$\beta_t = 2N_{c} g_t^{-2} \xi$.
Therefore, when the value for the beta-function is available, 
we can determine these anisotropy coefficients 
by measuring $r_t$ from the finite temperature transition 
curve in the $(\beta_s, \beta_t)$ plane.%
\footnote{
A similar approach was proposed in \cite{montvay}.
}

In order to determine the transition curve in the coupling 
parameter space, we compute the rotated Polyakov loop 
\begin{eqnarray}
L = z \, \frac{1}{N_s^3} \sum_{\vec{x}} \frac{1}{N_{c}} 
{\rm Tr} \prod_{t=1}^{N_t} U_4( \vec{x},t )
\end{eqnarray}
as a function of $(\beta_s, \beta_t)$, where $z$ is a $Z(N_c)$ phase 
factor ($z^{N_c} = 1$) such that $\arg(L) \in (-\pi/N_c,\pi/N_c]$.
We define the transition point as the peak position of 
the susceptibility 
$\chi = N_s^3(\langle L^2 \rangle - \langle L \rangle^2)$ 
in $\beta$ for each fixed $\gamma$.

We compute the coupling parameter dependence of $\chi$ in the 
$(\beta_s,\beta_t)$ plane 
by applying the spectral density method \cite{swendsen} 
extended to anisotropic lattices. 
This enables us to compute the anisotropy coefficients 
directly from simulations at $\xi \approx 1$ 
without introducing an interpolation Ansatz. 
Another good feature of the spectral density method is that 
the method works well even with data obtained only on isotropic lattices.
Therefore, 
we can use data from previous high statistic simulations 
performed on isotropic lattices,
when the time histories of the Polyakov loop and space-like and 
time-like plaquettes are available near the transition point.

Fitting the transition curve with a polynomial 
\begin{equation}
 \beta_{c} (\gamma) = \sum_{n = 0}^{n_{\rm max}} f_n \, (\gamma - 1)^{n},
\label{eq:polynomial}
\end{equation}
with $f_n$ the fitting parameters, 
the slope $ r_t$ 
is given by 
\begin{eqnarray}
r_t = 
\left(\frac{{\rm d} ( \beta_{c} / \gamma )}
{{\rm d} ( \beta_{c} \gamma )}\right)_{\xi = 1}  
= \frac{ ({\rm d} \beta_{c}/{\rm d} \gamma)_{\xi = 1} - \beta_{c}} 
{ ({\rm d} \beta_{c} / {\rm d} \gamma)_{\xi = 1} + \beta_{c}}, 
\end{eqnarray}
where $({\rm d} \beta_{c} / {\rm d} \gamma )_{\xi = 1} = f_{1}.$ 
The range of $\beta$ and $\gamma$ in which the spectral density method
is reliable is estimated by the condition that the statistical error 
for the reweighting factor 
(which is $\langle e^{-\Delta S} \rangle$ when the number of simulation 
points is one) is less than 0.5\%.
We confirm that the results are completely stable under a variation 
of $n_{\rm max}$ when we restrict ourselves to the range discussed above.
Choosing a range of $\gamma$ around 1 in such a way that 
the transition curve is almost straight, 
we use $n_{\rm max}=3$ for the final results.

\section{Results for $SU(2)$}
\label{sec:su2}

We first test the method for the case of $SU(2)$ gauge theory 
at the transition point $\beta_c$ for $N_t=4$ and 5. 
Although the method should work well with data only from isotropic 
lattices, in order to confirm it, we perform Monte Carlo simulations 
also on several anisotropic lattices for $SU(2)$.
On a $ 16^{3} \times 4 $ lattice, we perform simulations 
at $ (\beta_s, \beta_t ) = (2.300, 2.300)$, 
(2.302, 2.302), (2.296, 2.306), and (2.307, 2.298).
On a $20^3\times5$ lattice, we simulate at 
$(\beta_s, \beta_t)=(2.373,2.373)$, (2.375,2.375), (2.380,2.370)
and (2.368,2.378).
At each $(\beta_s, \beta_t)$ on the $N_t=4$ (5) lattice, 
we accumulate 500,000 (1,250,000) configurations, each separated by
10 heat-bath sweeps, after thermalization.
The statistical errors are estimated using the jackknife method 
with the bin size of 1000 configurations.
We confirm that the errors are stable under a wide variation 
of the bin size around this value.

\begin{figure}[tb]
\centerline{
\epsfxsize=0.8\textwidth \epsfbox{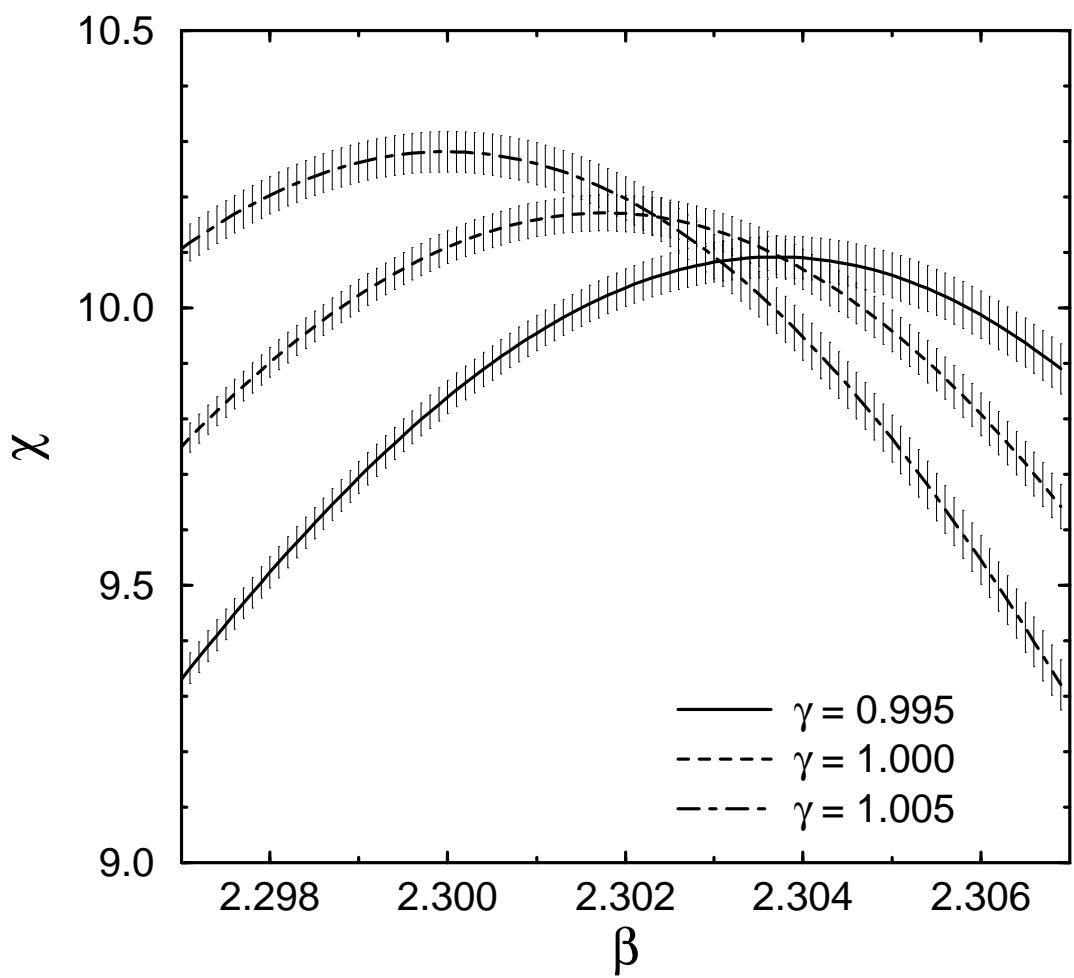}
}
\vspace{-3cm}
\caption{
Polyakov loop susceptibility in the $SU(2)$ gauge theory 
on a $16^3\times4$ lattice at $\gamma = 0.995$, 1.0 and 1.005.
Errors are estimated by a jackknife method.
}
\label{fig:su2sus}
\end{figure}

\begin{figure}[tb]
\vspace*{-1.0cm}
\centerline{
(a)\epsfxsize=9cm\epsfbox{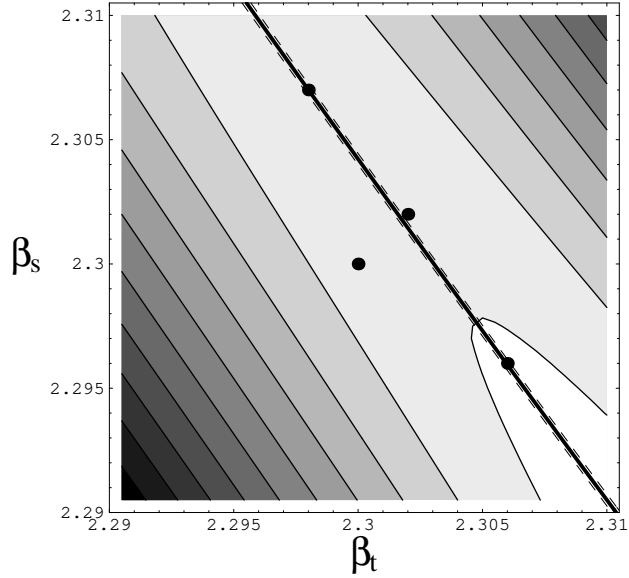}
\vspace*{-0.5cm}
}
\centerline{
(b)\epsfxsize=9cm\epsfbox{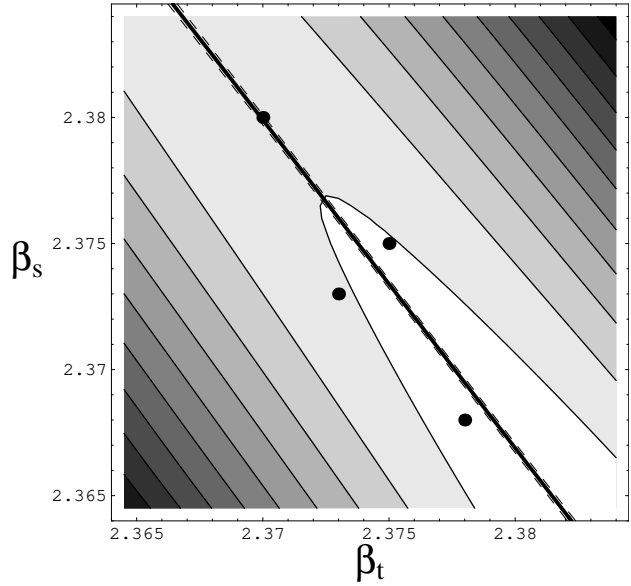}
}
\vspace{5mm}
\caption{
Polyakov loop susceptibility in the $SU(2)$ gauge theory 
as a function of $(\beta_s, \beta_t)$
obtained on (a) $16^3\times4$ and (b) $20^3\times5$ lattices.
Simulation points are shown by filled circles.
The bold lines represent the peak position of the susceptibility
and the dashed lines their errors.
The magnitude of the susceptibility is shown by tone
for the range (a) $8.2 < \chi < 10.4 $ and (b) $ 9.0 < \chi < 11.2$,
respectively, 
where different tone corresponds to a difference $\Delta\chi=0.2$.
}
\label{fig:su2bsbtp}
\end{figure}

Computing the susceptibility in the $(\beta_s,\beta_t)$ plane 
using data at each simulation point, we check that the results
agree well with each other, i.e.\ the results for the susceptibility 
from isotropic lattices coincide with the results from anisotropic lattices. 
For the rest of this section,
we combine the results for all four $(\beta_s,\beta_t)$ combinations
to compute the susceptibility with the spectral density method.
In Fig.\ref{fig:su2sus}, we plot the susceptibility for $N_t=4$ at 
$\gamma = 0.995$, 1.000, and 1.005.
The results for the peak position $\beta_c$ of the susceptibility 
computed at various values of $\gamma$ are summarized in 
Fig.~\ref{fig:su2bsbtp} for $N_t=4$ and 5.

Fitting the results for the transition curve, we obtain the values for 
$\beta_c$ and $r_t$ at $\xi=1$, as summarized in Table~\ref{tab:su2rt}. 
Combining the values of $r_t$ with a result of the $SU(2)$ 
beta-function\cite{engels} at $\beta_c(\xi\!=\!1)$, 
we obtain the anisotropy coefficients (\ref{eq:dgamdxi}) and (\ref{kc}).
The results are summarized in Table~\ref{tab:su2cst}. 
Because no errors for the beta-function are given in \cite{engels}, 
we disregard their contribution to the errors of the anisotropy coefficients.

\begin{figure}[tb]
\centerline{
\epsfxsize=0.8\textwidth \epsfbox{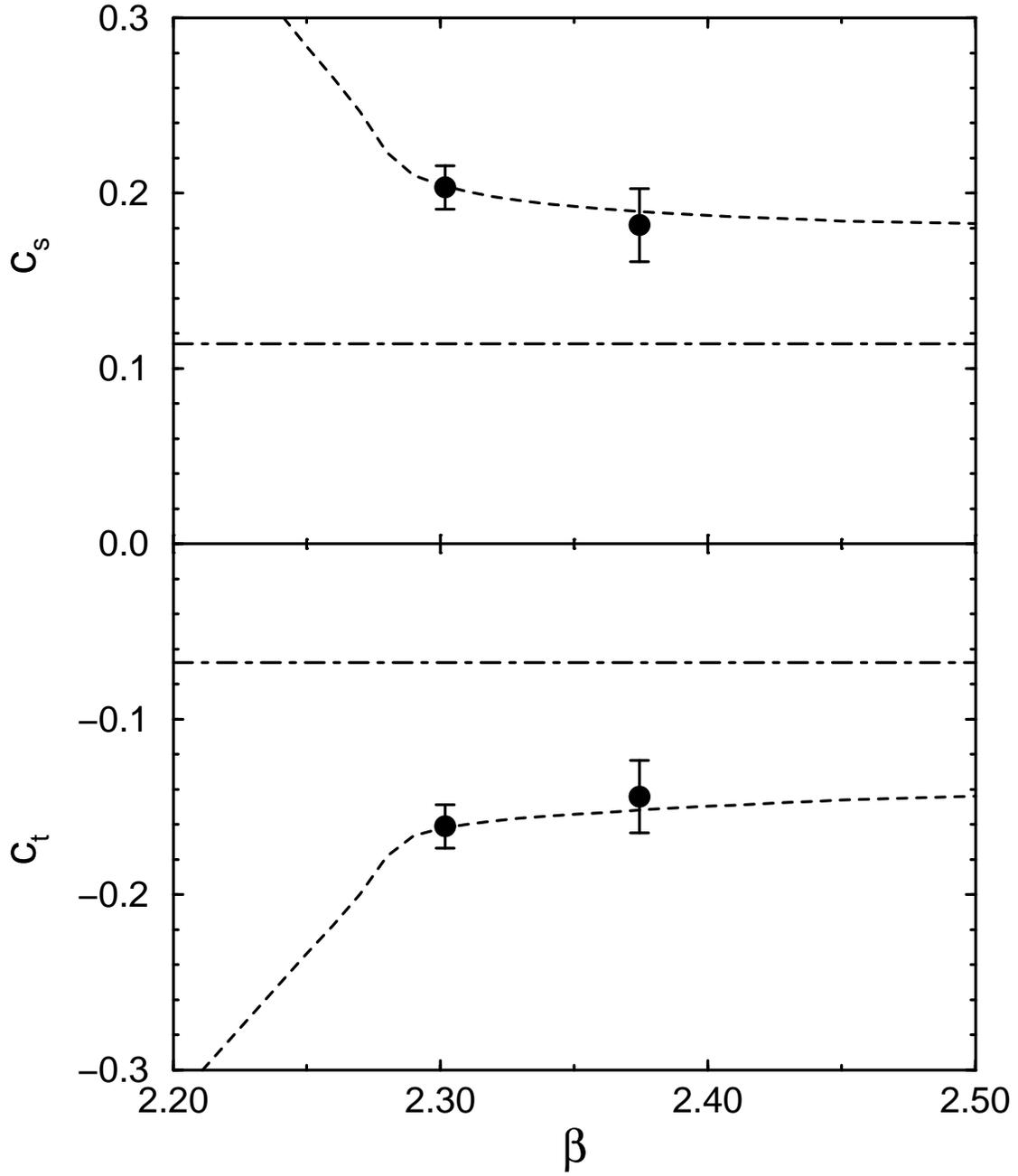}
}
\vspace{5mm}
\caption{
Anisotropy coefficients $c_s$ and $c_t$ for the $SU(2)$ gauge theory.
Our non-perturbative results are given by filled circles.
The dot-dashed curves are the results of the perturbation theory 
\protect\cite{karsch}.
The dotted curves are the results from the integral method
\protect\cite{engels}.
No errors are published for these curves.
}
\label{fig:su2kc}
\end{figure}

In Fig.~\ref{fig:su2kc}, we compare our results for the Karsch coefficients 
with the results of the perturbation theory (dot-dashed curves) 
\cite{karsch}
and the integral method (dotted curves) \cite{engels}.
We find significant discrepancies between our results 
and the results of the perturbation theory.
On the other hand,
our results are consistent with the results from the integral method.

\section{Results for $SU(3)$}
\label{sec:su3}

Let us now study the more realistic case of the $SU(3)$ gauge theory.
We analyze the high statistic data for the $SU(3)$ gauge theory
obtained by the QCDPAX Collaboration \cite{QCDPAX}.
Simulations were performed at the deconfining transition point for
$N_t=4$ and 6.
For $N_t=4$, the lattice sizes are $24^2\times 36\times 4$ and 
$12^3\times24\times4$, with $712\,000$ and $910\,000$
pseudo heat-bath iterations, respectively.
For $N_t=6$, data on $36^2 \times 48 \times 6$, $24^3\times6$, and
$20^3\times6$ lattices with $1\,112\,000$, $480\,000$, and
$376\,000$ iterations are available.
The Polyakov loop and the plaquettes are measured every iteration.
Details of the simulation parameters are given in \cite{QCDPAX}.
For the bin size in the jack-knife analysis, we adopt the 
same values as in \cite{QCDPAX}.

\subsection{Anisotropy coefficients}

\begin{figure}[tb]
(a)
\centerline{
\epsfxsize=9cm \epsfbox{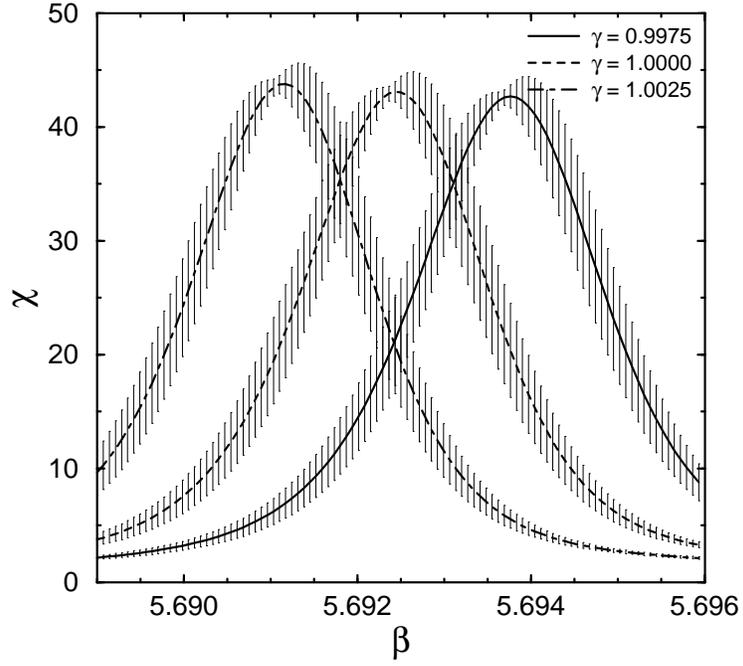}
}
\vspace*{-4cm}
(b)
\centerline{
\epsfxsize=9cm \epsfbox{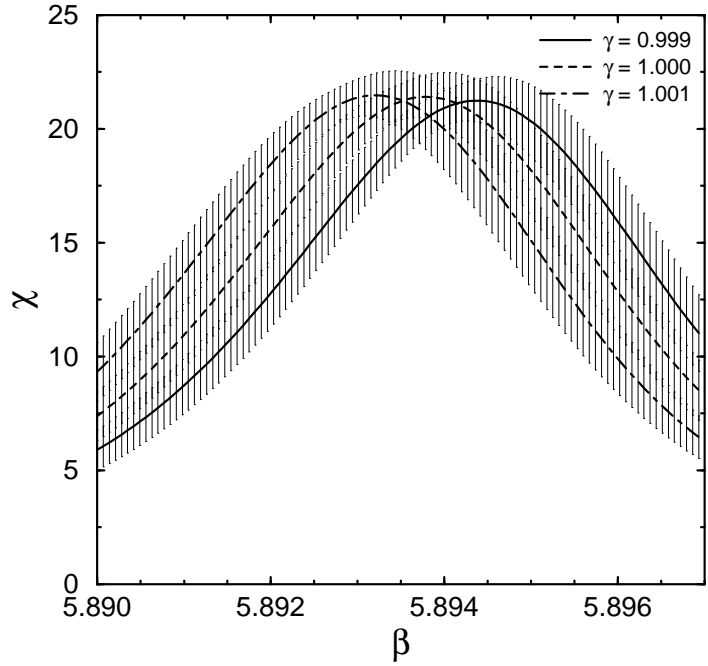}
}
\vspace*{-3cm}
\caption{
Polyakov loop susceptibility in the $SU(3)$ gauge theory 
obtained 
(a) on the $24^2\times36\times4$ lattice at 
$\gamma = 0.9975$, 1.0 and 1.0025,
and (b) on the $36^2\times48\times6$ lattice at 
$\gamma = 0.999$, 1.0 and 1.001.
}
\label{fig:su3sus}
\end{figure}

\begin{figure}[tb]
\vspace*{-1.0cm}
\centerline{
(a)\epsfxsize=9cm\epsfbox{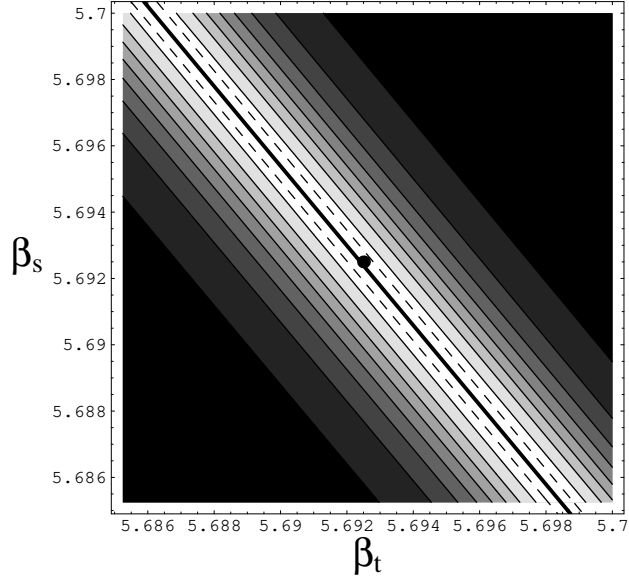}
\vspace*{-0.5cm}
}
\centerline{
(b)\epsfxsize=9cm\epsfbox{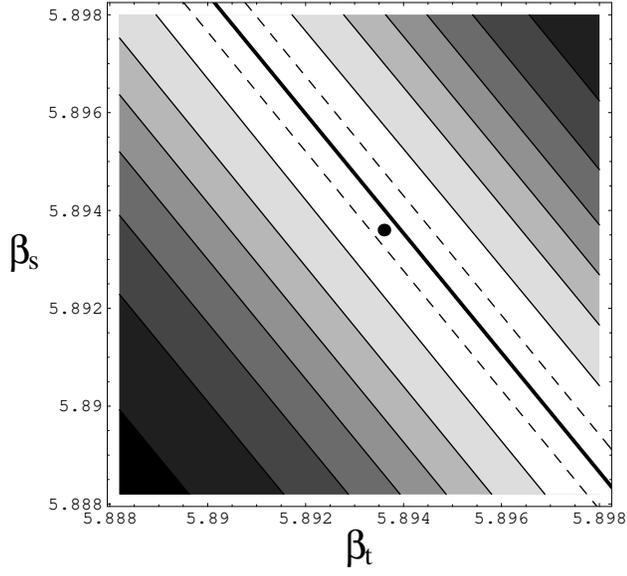}
}
\vspace{5mm}
\caption{
The same as Fig.~\protect\ref{fig:su2bsbtp} for the $SU(3)$ gauge theory
on (a) $24^3\times36\times4$ and (b) $36^2\times48\times6$ lattices.
The range of $\chi$ plotted and the width $\Delta\chi$ for a tone
are (a) 0.0 -- 45.0, 5.0 and (b) 2.5 -- 22.5, 2.5, respectively.
}
\label{fig:su3bsbtp}
\end{figure}

The results for the susceptibility on the largest spatial lattices
are given in Figs.~\ref{fig:su3sus} and \ref{fig:su3bsbtp}.
Because the transition is of first order for $SU(3)$, the peak of 
the susceptibility is quite clear when the spatial
lattice size is large enough, 
as shown in Figs.~\ref{fig:su3sus} and \ref{fig:su3bsbtp}.
(Note the difference in the vertical scales between Figs.~\ref{fig:su2sus} 
and \ref{fig:su3sus}.)

Our results for the slope $r_t$ are summarized in Table~\ref{tab:su3rt}.
Except for the case of the $24^3\times6$ lattice
where the simulation point is slightly off the transition point,
the errors become larger with decreasing spatial volume, 
because the peak of the susceptibility becomes less clear on small 
lattices.
From Table~\ref{tab:su3rt}, we find that the slopes at $N_t=4$ 
with different spatial lattice volumes completely agree with each other.
As shown in Figs.~\ref{fig:su3sus} and \ref{fig:su3bsbtp}, 
the peak of the susceptibility for
$N_t=6$ is less sharp compared with that for $N_t=4$ with the same 
relative spatial volume $(N_s/N_t)^3$ due to the fact that 
the transition is weaker for $N_t=6$ \cite{QCDPAX}. 
Therefore, with comparable statistics, $r_t$ has a larger statistical 
error for $N_t=6$.
Unlike in the case of $N_t=4$, the central values for the slope for $N_t=6$ 
given in Table~\ref{tab:su3rt} 
vary with the spatial volume by about one standard deviation. 
However, because the volume dependence is not uniform,
we consider that it is caused by statistical fluctuations.
We use the values obtained on the largest spatial lattices 
for our final results.

Our results for the anisotropy coefficients are summarized in 
Table~\ref{tab:su3cst}.
For our final results, we adopt the beta-function computed from a 
recent string tension data by the SCRI group \cite{edwards}.
See 
a subsection below 
for a discussion about the influence on the results from
the choice of the beta-function.

\subsection{Pressure gap and latent heat}

As an application of our non-perturbative anisotropy coefficients, 
we reanalyze the thermodynamic quantities $\epsilon$ and $p$ at the 
deconfining transition point using the plaquette data by the
QCDPAX Collaboration \cite{QCDPAX}.
In terms of the slope $r_t$ and the beta-function, 
the conventional combinations $\epsilon-3p$ and $\epsilon+p$
are given by 
\begin{eqnarray}
(\epsilon-3p)/T^4 &=& 
- 3 N_t^4 \, a \frac{{\rm d}\beta}{{\rm d}a} \,
\{\langle P_s \rangle + \langle P_t \rangle\ - 2\langle P \rangle_0\}, 
\label{eq:e3p} \\
(\epsilon+p)/T^4 &=& 
3 N_t^4 \, a \frac{{\rm d}\beta}{{\rm d}a} \, 
\frac{r_t-1}{r_t+1} \,
\{\langle P_s \rangle - \langle P_t \rangle\}.
\label{eq:emp}
\end{eqnarray}
At a first order transition point, we have a finite gap for energy 
density, the latent heat, but expect no gap for pressure.
It is known that the perturbative anisotropy coefficients have a
difficulty which leads to 
a non-vanishing pressure gap at the deconfining transition
point: $\Delta p / T^4 = -0.32(3)$ and $-0.14(2)$ at
$N_t=4$ and 6 \cite{QCDPAX}.

New values for the gaps in $\epsilon$ and $p$ using our non-perturbative 
anisotropy coefficients are summarized in Table~\ref{tab:su3ep}.
For the pressure gap, we obtain 
\begin{eqnarray}
\Delta p/T^4 = \left\{ 
  \begin{array}{rc} 0.001(15) & \; \mbox{for $N_t=4$,} \\
                   -0.003(17) & \; \mbox{for $N_t=6$.}
  \end{array} \right.
\end{eqnarray}
We find that the problem of non-zero pressure gap is completely
resolved with our non-perturbative anisotropy coefficients.

\subsection{Choice of the beta-function}
\label{ref:betafn}

In Table~\ref{tab:su3cst}, we study the influence of the choice
of the beta-function on the anisotropy coefficients. 
We compare 
(i) the beta-function computed from a recent string tension data 
by the SCRI group \cite{edwards},
(ii) that from a MCRG study by the QCDTARO Collaboration \cite{taro}, 
and (iii) that from a study of $\beta_c(N_t)$ by the Bielefeld group 
\cite{boyd}.
The SCRI beta-function is computed using a fit of the string tension 
for $5.6 \leq \beta \leq 6.5$.
We note that the QCDTARO beta-function is based on a fit of mean-field 
improved gauge coupling constant using the results of plaquette at 
$\beta > 5.8$; i.e.\ $\beta_c(N_t\!=\!4) \approx 5.69$ is slightly off 
the range of validity \cite{taro,Miyamura}.
Also the beta-function by the Bielefeld group seems to be problematic
around $\beta_c(N_t\!=\!4)$, because it is largely affected by the data
of $\beta_c(N_t\!=\!3)$ where we cannot expect universal scaling.
Accordingly, the beta-function of the Bielefeld group shows a systematic 
deviation from the data of a MCRG study at 
$\beta$ {\raise0.3ex\hbox{$<$\kern-0.75em\raise-1.1ex\hbox{$\sim$}}} 
6 \cite{boyd}.

\begin{figure}[tb]
\centerline{
\epsfxsize=0.8\textwidth \epsfbox{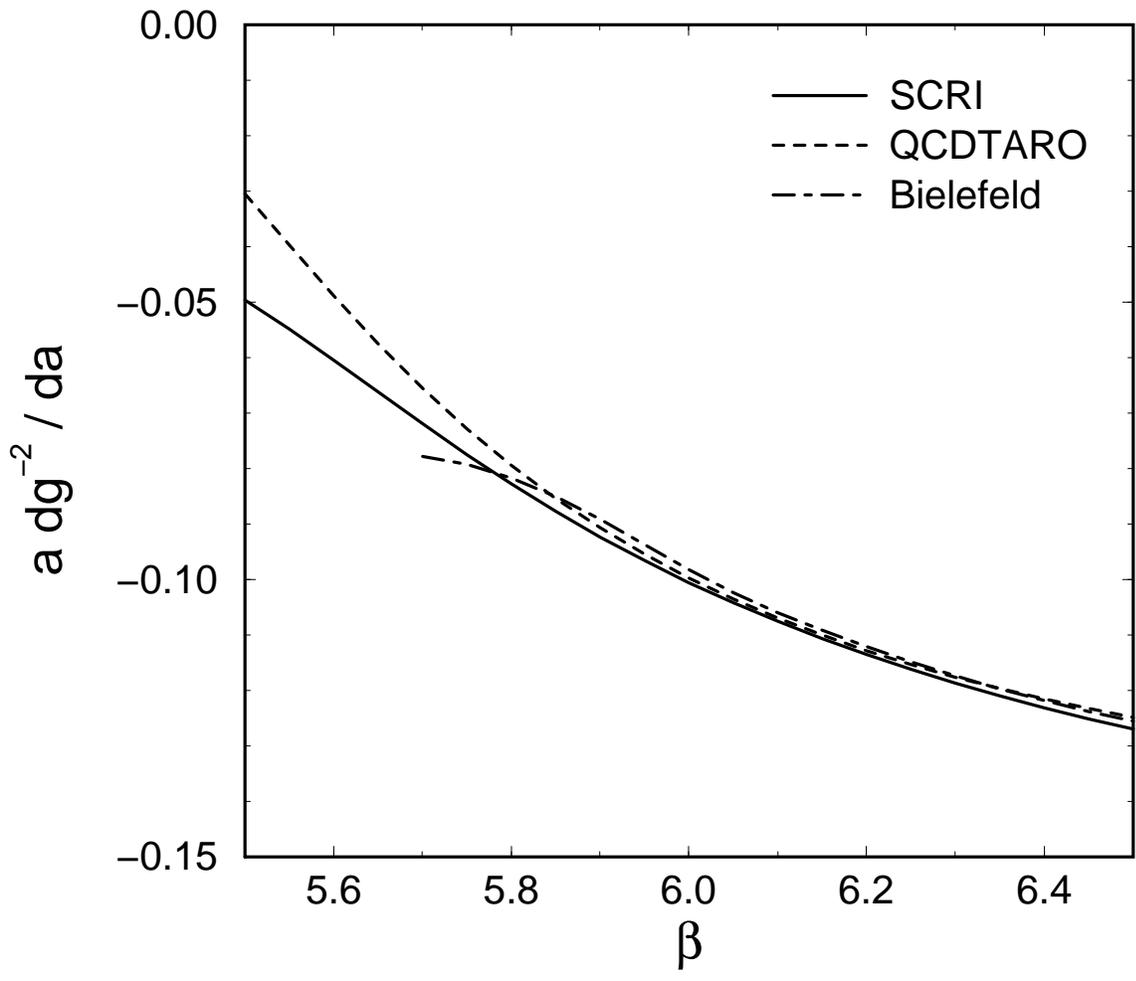}
}
\vspace{-3cm}
\caption{
Non-perturbative beta-functions in the $SU(3)$ gauge theory.
}
\label{fig:su3beta}
\end{figure}

These beta-functions are plotted in Fig.~\ref{fig:su3beta}.
At $\beta_c(N_t\!=\!6)$, different beta-functions coincide 
with each other within 5\%,
while, at $\beta_c(N_t\!=\!4)$, they vary by about 20\%.
Because only the SCRI beta-function is reliable at $\beta_c(N_t\!=\!4)$
as discussed in the previous paragraph, 
we adopt the SCRI beta-function for our final results.

In order to compare the anisotropy coefficients from different 
references, however, it is important to check the effect of 
the beta-function on the results.
From Table~\ref{tab:su3cst},
we see that the results for the anisotropy coefficients 
using different beta-functions 
agree well with each other at $N_t=6$.
At $N_t=4$, however, the anisotropy coefficients depend very much 
on the choice of the beta-function. 
Accordingly, we find that the results for the latent heat
are consistent with each other at $N_t=6$:
$\Delta\epsilon/T^4= 1.569(40)$, 1.539(39), and 1.515(38)
with SCRI, QCDTARO, and Bielefeld beta-functions, respectively.
At $N_t=4$, we find a sizable dependence on the choice of the beta-function:
$\Delta\epsilon/T^4= 2.074(34)$, 1.877(30), and 2.265(37) 
using SCRI, QCDTARO, and Bielefeld beta-functions. 
For the pressure gap, on the other hand,
because the beta-function appears only as a common overall factor
in (\ref{eq:e3p}) and (\ref{eq:emp}), 
the conclusion that $\Delta p$ vanishes with our anisotropy coefficients
does not depend on the choice of the beta-function.

\subsection{Comparison with other methods}

In Fig.~\ref{fig:su3kc}, we summarize our results for the Karsch
coefficients together with previous values; 
the perturbative results \cite{karsch},
results from the integral method \cite{boyd},
and those from the matching of Wilson loops on anisotropic lattices
\cite{scheideler,klassen}.
No errors are published for the results from the integral method.
We find that all non-perturbative methods give values 
which deviate from the results in the perturbation theory. 

Comparing the results from different non-perturbative methods, 
we find that, 
although the deviations from the perturbation theory are 
roughly consistent with each other, 
the central values are different by more than three standard deviations,
when we take the published errors.

We think that one origin of the variation among different methods at 
$\beta_c(N_t\!=\!4)$ is the beta-function.
Note that the results from Refs.~\cite{scheideler} (matching method) 
and \cite{boyd} (integral method) are computed using the beta-function 
of the Bielefeld group, while our results and the results from 
Ref.~\cite{klassen} (matching method) are using the SCRI beta-function.
From Table~\ref{tab:su3cst}, we note that, if we adopt the beta-function
of the Bielefeld group, our results are consistent with those 
of Ref.~\cite{scheideler} at $\beta_c(N_t\!=\!4)$.

At $\beta_c(N_t\!=\!6)$, on the other hand, 
the difference in the results is not due to the beta-function, 
because the systematic error due to the choice of the beta-function is small
as discussed in the previous subsection.
In order to see this, we study $\partial\gamma/\partial\xi$,
which can be computed without using the beta-function in the matching method.
The values of $\partial\gamma/\partial\xi$ obtained in Ref.~\cite{klassen} 
are reported to be consistent 
with those from the integral method \cite{boyd}, 
but are different to another result from the matching method 
\cite{scheideler}.
Performing a quadratic interpolation in $\beta$, 
we find $\partial\gamma/\partial\xi \simeq 0.64(1)$ \cite{klassen}, 
$0.66(2)$ \cite{boyd}, and $0.74(2)$ \cite{scheideler} 
at $\beta_c(N_t\!=\!6)$.
Our result $0.707(10)$ given in Table~\ref{tab:su3cst} 
is around the center of these values.
A careful study of systematic errors in each method is required to 
understand the variation between different methods.

\section{Conclusions}
\label{sec:conclusions}

We have computed the anisotropy coefficients 
for the $SU(2)$ and $SU(3)$ gauge theories 
by measuring the transition curve of the deconfining transition in 
the $(\beta_s, \beta_t)$ plane.
One of the essential ingredients of our approach is the application of 
the spectral density method, that enables us to determine the
anisotropy coefficients directly from simulations at $\xi \approx 1$.
We note that the spectral density method is useful to avoid 
interpolation Ans\"atze also in the matching method.

\begin{figure}[tb]
\centerline{
\epsfxsize=0.8\textwidth \epsfbox{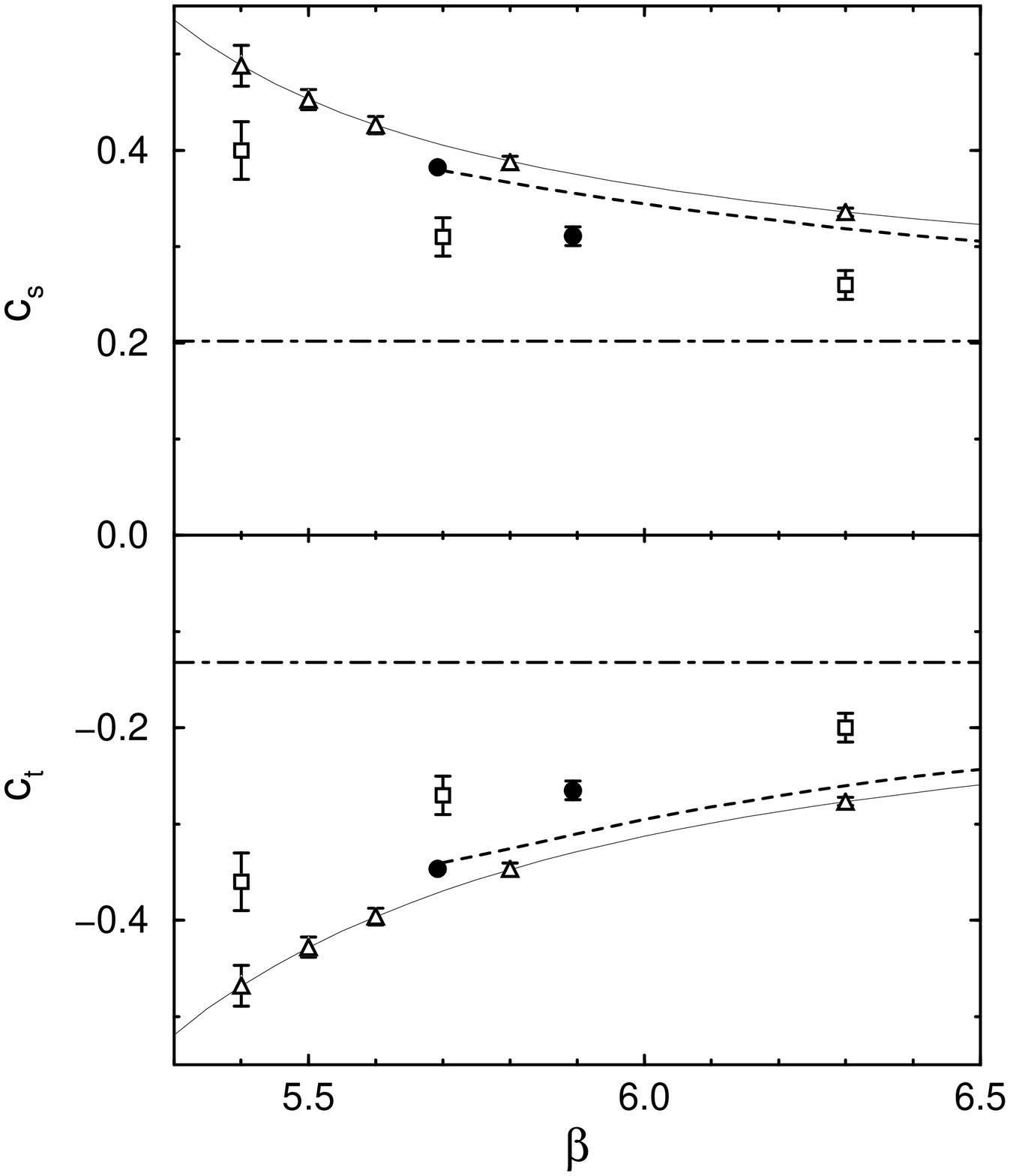}
}
\vspace{5mm}
\caption{
Anisotropy coefficients in the $SU(3)$ gauge theory.
Our non-perturbative results are given by filled circles.
The dot-dashed curves are the results of the perturbation theory 
\protect\cite{karsch}.
The open squares are those from a matching of Wilson loops 
\protect\cite{scheideler}.
Open triangles and thin lines are the results of a matching method 
\protect\cite{klassen} combined with the SCRI beta-function 
\protect\cite{edwards}.
The dotted curves are the results from the integral method
\protect\cite{boyd}.
No errors are published for the results from the integral method.
}
\label{fig:su3kc}
\end{figure}

Our non-perturbative results for the anisotropy coefficients are 
summarized in Tables~\ref{tab:su2cst} and \ref{tab:su3cst}.
Our results shown in Fig.~\ref{fig:su3kc} suggest that 
the Karsch coefficients converge to the perturbative values
slightly faster than that suggested by the central values from 
Refs.~\cite{boyd} and \cite{klassen}.
Applying the results for $SU(3)$, we reanalyzed the thermodynamic 
quantities at the deconfining transition point on $N_t=4$ and 6 lattices.
We obtain vanishing pressure gaps with our non-perturbative anisotropy 
coefficients, thereby solving
a longstanding problem of non-zero pressure gap 
with the perturbative coefficients.

\vspace{5mm}

We are grateful to O.\ Miyamura, A.\ Nakamura and H.\ Matsufuru 
for useful discussions and sending us the data for the QCDTARO 
beta-function.
We also thank A.\ Ukawa, T.\ Yoshi\'e, Y.\ Aoki, T.\ Kaneko, 
R.\ Burkhalter and H.P.\ Shanahan for helpful suggestions and comments. 
This work is in part supported by 
the Grants-in-Aid of Ministry of Education,
Science and Culture (Nos.~08NP0101 and 09304029).
SE is supported by the Japan Society for the Promotion of Science.

\begin{table}[tb]
\caption{Results for $\beta_c$ and the slopes at $\xi=1$ in the 
$SU(2)$ gauge theory.
The column ``$\gamma$-range'' is for the range of $\gamma$
used in the fit for the slope $d\beta_c/d\gamma$.
}
\label{tab:su2rt}
\begin{center}
\begin{tabular}{ccccc}
\hline
lattice & $\beta_c$ & $\gamma$-range & $d\beta_c/d\gamma$ & $r_t$ \\
\hline
$16^3\times4$ & 2.30177(9) & 0.995 -- 1.005 & $-$0.370(12) & $-$1.384(14) \\
$20^3\times5$ & 2.37430(8) & 0.995 -- 1.005 & $-$0.312(15) & $-$1.303(17) \\
\hline
\end{tabular}
\end{center}
\end{table}

\begin{table}[tb]
\caption{$SU(2)$ anisotropy coefficients at $\xi=1$ 
using the beta-function $a{\rm d}g^{-2}/{\rm d}a$ 
obtained by the Bielefeld group \protect\cite{engels}.
}
\label{tab:su2cst}
\begin{center}
\begin{tabular}{ccccc}
\hline
lattice & $\partial\gamma/\partial\xi$ & $c_s$ & $c_t$ 
& $a{\rm d}g^{-2}/{\rm d}a$ \\
\hline
$16^3\times4$ & 0.683(21) & 0.203(12) & $-$0.161(12) & $-$0.08439 \\
$20^3\times5$ & 0.725(35) & 0.182(21) & $-$0.144(21) & $-$0.07544 \\
\hline
\end{tabular}
\end{center}
\end{table}

\begin{table}[tb]
\caption{The same as Table~\protect\ref{tab:su2rt} for $SU(3)$ 
using the data by the QCDPAX Collaboration \protect\cite{QCDPAX}.
}
\label{tab:su3rt}
\begin{center}
\begin{tabular}{ccccc}
\hline
lattice & $\beta_c$ & $\gamma$-range & $d\beta_c/d\gamma$ & $r_t$ \\
\hline
$24^2\times36\times4$ & 5.69245(23) & 0.9975 -- 1.0025 & 
$-$0.5193(23) & $-$1.2008(10) \\ 
$12^2\times24\times4$ & 5.69149(42) & 0.995 -- 1.005 & 
$-$0.5183(52) & $-$1.2004(22) \\
\hline
$36^2\times48\times6$ & 5.89379(34) & 0.999 -- 1.001 & 
$-$0.5844(83) & $-$1.2201(35) \\
$24^3\times6$         & 5.89292(87) & 0.999 -- 1.001 & 
$-$0.542(33) & $-$1.202(14) \\
$20^3\times6$         & 5.8924(14)  & 0.9975 -- 1.0025 & 
$-$0.622(34) & $-$1.236(14) \\
\hline
\end{tabular}
\end{center}
\end{table}

\begin{table}[tb]
\caption{$SU(3)$ anisotropy coefficients at $\xi=1$,
using the values for the beta-function $a{\rm d}g^{-2}/{\rm d}a$
by the SCRI group\protect\cite{edwards}, 
the QCDTARO Collaboration\protect\cite{taro}, 
and the Bielefeld group\protect\cite{boyd}.
For our final results, we take the values obtained on the largest 
spatial lattices using the SCRI beta-function.
Because the errors for the beta-function are not given in the papers, 
we disregard their contribution to the errors of the anisotropy 
coefficients in this table.
See text for details.
}
\label{tab:su3cst}
\begin{center}
\begin{tabular}{ccccl}
\hline
lattice & $\partial\gamma/\partial\xi$ & $c_s$ & $c_t$ 
& \hspace{10mm} $a{\rm d}g^{-2}/{\rm d}a$ \\
\hline
$24^2\times36\times4$ & 0.6159(27) & 0.3822(26) & $-$0.3466(26) & $-$0.07108 \hspace{3mm} SCRI      \\
                      & 0.5575(25) & 0.4359(23) & $-$0.4037(23) & $-$0.06434 \hspace{3mm} QCDTARO   \\
                      & 0.6728(30) & 0.3299(28) & $-$0.2910(28) & $-$0.07764 \hspace{3mm} Bielefeld \\
$12^2\times24\times4$ & 0.6161(62) & 0.3819(59) & $-$0.3464(59) & $-$0.07097 \hspace{3mm} SCRI      \\
                      & 0.5573(56) & 0.4360(53) & $-$0.4039(53) & $-$0.06418 \hspace{3mm} QCDTARO   \\
                      & 0.6738(68) & 0.3288(64) & $-$0.2900(64) & $-$0.07761 \hspace{3mm} Bielefeld \\
\hline                          
$36^2\times48\times6$ & 0.7068(100)& 0.3109(98) & $-$0.2650(98) & $-$0.09179 \hspace{3mm} SCRI      \\
                      & 0.6936(98) & 0.3235(96) & $-$0.2784(96) & $-$0.09008 \hspace{3mm} QCDTARO   \\
                      & 0.6826(96) & 0.3340(95) & $-$0.2897(95) & $-$0.08864 \hspace{3mm} Bielefeld \\
$24^3\times6$         & 0.762(47)  & 0.257(46)  & $-$0.211(46)  & $-$0.09172 \hspace{3mm} SCRI      \\
                      & 0.747(46)  & 0.271(45)  & $-$0.226(45)  & $-$0.08999 \hspace{3mm} QCDTARO   \\
                      & 0.736(45)  & 0.282(45)  & $-$0.237(45)  & $-$0.08857 \hspace{3mm} Bielefeld \\
$20^3\times6$         & 0.663(36)  & 0.354(35)  & $-$0.308(35)  & $-$0.09167 \hspace{3mm} SCRI      \\
                      & 0.651(35)  & 0.366(35)  & $-$0.321(35)  & $-$0.08994 \hspace{3mm} QCDTARO   \\
                      & 0.640(35)  & 0.375(34)  & $-$0.331(34)  & $-$0.08853 \hspace{3mm} Bielefeld \\
\hline
\end{tabular}
\end{center}
\end{table}

\begin{table}[tb]
\caption{Gaps for thermodynamic quantities in the $SU(3)$ gauge theory
at the deconfining transition point using our non-perturbative
anisotropy coefficients.
Plaquette data are taken from Ref.~\protect\cite{QCDPAX}.
The low temperature hadronic phase (had) and the high temperature 
quark-gluon-plasma phase (QGP) are separated as described
in \protect\cite{QCDPAX}.
We reanalyze $(\epsilon-3p)/T^4$ also, using the SCRI beta-function. 
}
\label{tab:su3ep}
\begin{center}
\begin{tabular}{lcc}
\hline
lattice            & $24^2\times36\times4$ & $36^2\times48\times6$ \\
$\beta$                       &   5.6925    &    5.8936    \\
\hline
$\Delta(\epsilon+p)/T^4$      &   2.075(42) &    1.565(51) \\
$\Delta(\epsilon-3p)/T^4$     &   2.072(43) &    1.578(42) \\
$\Delta \epsilon/T^4$         &   2.074(34) &    1.569(40) \\
$\Delta p/T^4$                &   0.001(15) & $-$0.003(17) \\
\hline
\end{tabular}
\end{center}
\end{table}

\end{document}